\documentclass{article}
 \usepackage[pdftex]{graphicx}
 \usepackage{amsmath,amssymb}
 \usepackage{booktabs, rotating, tabularx, url}
 \usepackage{todonotes}
 \usepackage{algorithm}
 \usepackage{algpseudocode}
 \usepackage{multirow}
 \usepackage{color}

\usepackage{tikz,pgfplots,pgfplotstable} 
\pgfplotsset{compat=newest}
\usetikzlibrary{calc,3d,arrows,patterns,decorations}
\tikzset{>=stealth'}
\usetikzlibrary{positioning} 
\usepackage[abs]{overpic}
  
\date{\today}
 
\title{Resilience for Multigrid Software at the Extreme Scale}
   
\newcommand{\commentGeneric}[2]{~\\\fbox{\begin{minipage}{0.9\columnwidth}
 \textbf{Kommentar #1:}\\[1ex]
 \begin{minipage}{0.975\textwidth}
 \it #2
 \end{minipage}\end{minipage}}\quad\\[1ex]}
\newboolean{SHOWCOMMENTS}
\setboolean{SHOWCOMMENTS}{false}
\ifthenelse{\boolean{SHOWCOMMENTS}}{%
\newcommand{\commentG}[1]{\commentGeneric{Gmeiner}{#1}}
\newcommand{\commentH}[1]{\commentGeneric{Huber}{#1}}

\newcommand{\commentW}[1]{\commentGeneric{Wohlmuth}{#1}}

}

{%
\def\commentG#1{}
\def\commentH#1{}
\def\commentR#1{}
\def\commentW#1{}
\def\todo#1{}
}

\def \super {\mbox{\footnotesize super}}
\def \ul {\underline}
\begin{document}

\author{Markus Huber$^{1}$ \and Bj\"orn Gmeiner$^{2}$ \and Ulrich R\"ude$^{2}$ \and Barbara Wohlmuth$^{1}$}
\footnotetext[1]{Institute for Numerical Mathematics (M2), Technische Universit{\"a}t M{\"u}nchen, Boltzmannstrasse~3, D--85748 Garching b. M\"unchen, Germany}
\footnotetext[2]{Department of Computer Science 10, FAU Erlangen-N{\"u}rnberg, Cauerstra\ss{}e 6, D--91058 Erlangen, Germany}

\maketitle

\begin{abstract}
Fault tolerant algorithms for the numerical approximation of elliptic partial
differential equations on modern supercomputers play a more and more important role in the future design of exa-scale enabled iterative solvers.
Here, we combine domain partitioning with
highly scalable geometric multigrid schemes to obtain fast and fault-robust
solvers in three dimensions.
The recovery strategy is based on a hierarchical hybrid concept where
the values on lower dimensional primitives such as faces are stored redundantly and thus can 
be recovered easily in case of a failure. The lost volume 
unknowns in the faulty region are re-computed approximately with multigrid cycles
by solving a local Dirichlet problem on the
faulty subdomain.
Different strategies are compared and evaluated with respect to
performance, computational cost, and speed up.
Especially effective are strategies  
in which 
the local recovery in the faulty region is
executed in parallel with global solves
and when the local recovery is additionally accelerated. 
This results in an asynchronous multigrid iteration
that can fully compensate faults.
Excellent parallel performance on a current peta-scale system is demonstrated.

	\vspace{2em}
	\noindent
	{\bf AMS: }65N55, 65Y05, 68Q85
\end{abstract}
%
%

 \pagestyle{myheadings}
 \thispagestyle{plain}
 \markboth{M. Huber, B. Gmeiner, U. R\"ude, B. Wohlmuth}{Resilience for Multigrid Software}

\section{Introduction}\label{sec:intro}   
%
Future high performance systems will be characterized by millions of
compute nodes that are executing up to a billion of parallel threads.
This compute power will be extremely expensive not only with respect
to acquisition costs but also due to
the operational costs,
whereby the energy consumption 
is becoming a major concern.
The increasing system size 
results in a higher probability  of failure of the components of the HPC-system \cite{Cappello09},
and thus fail-safe performance is one of the new challenges in 
extreme scale computing.

Faults can be classified in fail-stop and fail-continue, also called
hard and soft errors, respectively, see, e.g.,  \cite{BFHH12,CGGKS09}.
In the first case, the process stops, e.g.,
due to a permanent node crash or incorrect execution path
which interrupts the program and results in a loss of
the state of the process.
In the case of soft errors, the process continues but the
failure affects the execution through ``bit-flips'' (transient errors).
Fault tolerance  techniques can be categorized in 
hardware-based fault tolerance (HBFT) 
\cite{MKFM13,mukherjee2005},
system software-based fault tolerance (SBFT) 
\cite{bland2013post,BDBHBD13,BBBCD13,SBBC14,Zheng06},
and algorithm-based fault tolerance (ABFT) \cite{BFHH12,CSBS12,DJLMMP14,HA84,Malkowski10}.
For a general overview and a classification, we refer to \cite{Cappello09,CGGKS09,CGKKS14}. 

Achieving resilience is costly, since it always requires some form of redundancy,
and thus a duplication of system resources and extra energy.
In particular, traditional checkpoint strategies 
must collect and transfer the data regularly from 
all compute nodes and store the data to backup memory \cite{daly2003model,hursey2010coordinated,moody2010design,shahzad2013evaluation}. 
In  large systems, this may be too expensive and slow.
Consequently, algorithmic alternatives are required.
It is only natural  that the most efficient resilience techniques will have to
exploit specific features of the algorithms.

Under the assumption that the failure can be  
detected, such ABFT strategies implement the resilience in the algorithm itself
and thus guarantee reliable results. Originally, ABFT was proposed
by Huang and Abraham \cite{HA84} for
systolic arrays where checksums monitor the  data and are used for a
reconstruction. Later on, it was extended to applications in linear algebra
such as addition, matrix operations, scalar product, LU-decomposition, transposition and in fast Fourier transformation 
\cite{AnfinsonLuk88,BoleyBrentGolubLuk92,ChenDongarra08,LukPark88}.
The work by Davies and Chen \cite{DaviesChen13}  deals with
fault detection and correction during the calculation for dense matrix operations.
For sparse matrix iterative  solvers,  
such as SOR, GMRES, CG-iterations
the previously mentioned approaches
are not suitable due to a possible  high overhead 
\cite{SKB12} and were  extended by \cite{agullo13,BFHH12,Chen2013,RoyBanerjee93,Stoyanov_Webster_2013}. In 
\cite{cui_error-resilient_2013}, 
Cui et al. exploit the structure of a parallel subspace correction method such that the subspaces
are redundant on different processors, and the workload is efficiently balanced.

At this time, the immediate detection and the replacement of the faulty
components is not part of the standard message passing interface (MPI),
however, fault tolerant MPI versions, such as Harness FT-MPI 
\cite{fagg2000ft} or ULFM \cite{bland2013post,bland2012evaluation}
are under development. 
They do 
also not yet provide an instant reporting of the failure,
but may do so in future versions when the hardware itself is extended
to better support such features. For the purposes of this paper,
we assume that such extensions are already available.

In this article, we thus consider fail-stop errors as they
may occur in iterative schemes when solving
discretized elliptic partial differential equations.
We focus on the design of fault tolerant parallel geometric multigrid methods
since geometric multigrid methods are well-known for their 
asymptotic optimal complexity and excellent parallel efficiency 
\cite{brandt2011multigrid,chow2006survey,GRSWW13,GmeinerTextbook2015,
hackbusch1985multi,mcbryan1991multigrid,Sundar12}.
The influence of soft faults on an algebraic multigrid solver was studied
in \cite{CSBS12}.
Similar to \cite{cui_error-resilient_2013}, we pursue fault tolerance strategies  
for hard faults that
\begin{itemize}
 \item converge when a fault occurs, assuming it is detect\-able,
 \item minimize the delay in the solution process,
 \item minimize computational and communication overhead.
\end{itemize}

In order to compensate for a hard fault, we decouple the faulty
part from the intact part and exploit the fact that only relatively
little of the data must be stored redundantly so that a special 
recovery process can reconstruct the bulk of the missing data
efficiently.
Furthermore, we find that the redundant data as it is required for the reconstruction
of lost data is readily available in suitably designed data structures.
These data structures can be found in many
distributed memory parallel multigrid solvers that use a domain partitioning and ghost nodes
for communicating.

Our design goal is to minimize the time delay in an iterative solver
that seems inevitable when a fault has destroyed part of its internal state.
Here, however, we will show
that the  
flexibility of modern heterogeneous architectures 
can be used beneficially. 
If the recovery is executed with powerful enough resources, 
the loss can in many cases be compensated completely
and the time to solution experiences no delay.
In other words, by using a {\em computational superman} to recover the lost state,
we  are able to fully compensate a fault.

The rest of our paper is organized as follows: In Sec.\ \ref{sec:faultsolprocess},
we describe the model PDE and fault setting.
The required data structure  for our recovery strategies will be discussed in Sec. \ref{sec:HHG}.
The redundancy in the ghost layer data structures 
allows to combine full multigrid efficiency with
tearing and interconnecting strategies.  
  In Sec. \ref{sec:localrecovery}, we then develop  
local recovery strategies and study numerically their influence
on the global convergence. 
The main algorithmic result can
be found in  Sec. \ref{sec:globalrecovery}.
Here Dirichlet--Neumann and Dirichlet--Dirichlet coupling strategies are combined with
fast hierarchical multigrid methods.
The effect of the size of the faulty domain is illustrated for large scale computations.  
In Sec.~\ref{sec:parallelrecovery}, 
we test both new  global recovery strategies  on a 
state-of-the-art peta-scale system. 
To fully compensate for the fault with respect to the time to solution, we enhance the massively parallel geometric multigrid method 
by asynchronous components on partitioned domains.

\section{Model problem and fault setting}\label{sec:faultsolprocess}
In this section, we introduce the model PDE scenario, the notation for the geometrical multigrid solver   
and the  fault model 
for which we design our recovery techniques.

\subsection{Model problem and parallel multigrid solver}\label{sec:ModelProblem}
For the sake of simplicity, we will
illustrate the methods for  
 the Laplace equation in 3D with inhomogeneous Dirichlet boundary conditions
\begin{equation}\label{eq:Poisson}
-\Delta u = 0 ~~\mbox{in } \Omega, \qquad u = g ~~ \mbox{on } \partial \Omega 
\end{equation}
as PDE model problem. The generalization to more general boundary
conditions, right hand sides or other scalar elliptic problems is straightforward.
In Sec. \ref{sec:localrecovery}, we will also present a numerical example for the Stokes system.
Here,  $\Omega \subset \mathbb{R}^3$ is a bounded polyhedral domain
which is triangulated with an unstructured  
{\em base mesh}
$\mathcal{T}_{-2}$.
In the following, we will assume for simplicity that tetrahedral elements are used, but again
our  techniques generalize readily to hexahedral and hybrid meshes.
From this initial coarse mesh, a hierarchy of
meshes $\mathcal{T} := \{ \mathcal{T}_{l},~l=0, \ldots, L\}$ 
is constructed by successive uniform refinement, see, e.g., \cite{Bey:1995fk}.
This hierarchy is used 
to construct the geometric multigrid solver;
the choice to start with $l=0$ guarantees that there is at least one inner
degree of freedom if the mesh on level $l$ is restricted to one $T \in
\mathcal{T}_{-2}$. 

The discretization of \eqref{eq:Poisson}   
uses conforming linear finite elements (FE) on $\mathcal{T}_{l}$ 
that leads canonically to a nested sequence of finite 
element spaces $V_0 \subset V_1 \subset \ldots  \subset V_L \subset H^1(\Omega)$
and a corresponding family of  linear
systems 
\begin{equation}\label{eq:PoissonLin}
    A_l \ul{u}_l = \ul{f}_l;  ~ l=0, \ldots ,L,
\end{equation}
where the Dirichlet boundary conditions are included.
We apply multigrid correction schemes in V-cycles with standard components to \eqref{eq:PoissonLin}.
In particular, we use linear interpolation and its adjoint operator for the inter-grid transfer,
a hybrid variant of a Gauss-Seidel updating scheme in three pre- and post-smoothing  steps
and a preconditioned conjugate gradient (PCG) method as coarse grid solver.
For a general overview of multigrid methods, 
we refer to \cite{brandt2011multigrid,hackbusch1985multi}.
In highly parallel multigrid frameworks for distributed memory architectures \cite{balay2014petsc,bastian2008generic,falgout2002hypre,GRSWW13,Sundar12},
the work load for solving PDE problems is typically distributed to different
processes based on a geometrical domain partitioning. 
Here, the {\em base tetrahedral mesh} $\mathcal{T}_{-2}$ defines the partitioning
used for parallelization.
Each tetrahedron  $T \in \mathcal{T}_{-2}$ is  
associated with a processor, and this also
induces a process assignment for all refined meshes. 
In general,  
several of the coarse mesh tetrahedra would be assigned
to each processor so that a good load balancing occurs. 
In our simple model case, all tetrahedra are equally refined
and thus induce the same load, so that
for the sake of simplicity, we 
assume that the number of 
processors is equal to the  number  of tetrahedra in 
$\mathcal{T}_{-2}$, i.e.,  we have a one-to-one mapping
of base mesh tetrahedra to processors.
 
\subsection{Fault model and pollution effect}\label{sec:FaultModel}
For our study, we concentrate on a specific fault model
under assumptions similar to \cite{cui_error-resilient_2013,Harding14}.
We restrict our study, for simplicity, to the case that only one processor crashes.
All strategies can be 
extended easily to a failure of more processors, since
they only rely on the locality of the fault, see also Sec. \ref{sec:parallelrecovery},
where a large scale simulation 
with two faults at different locations
will be presented.

\begin{figure}[htb]
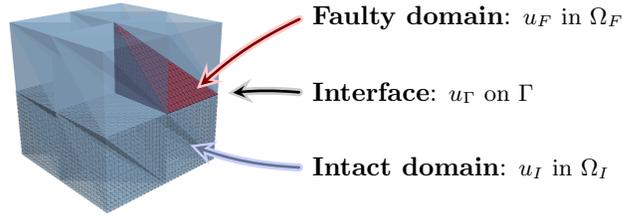

  \vspace*{-0.5cm}  \include{FaultDomain}
   \vspace{-0.5cm}
   \caption{Domain $\Omega$
	 with a faulty subdomain $\Omega_F$, interface $\Gamma$ and intact subdomain $\Omega_I$.}
   \label{fig:FaultTet}
\end{figure}

The {\em faulty} subdomain
$\Omega_F \subset \Omega$ is assumed to be
a single open tetrahedron in $\mathcal{T}_{-2}$, and $ \Omega_I := \Omega \setminus \overline \Omega_F$
is called  {\em intact} or {\em healthy} subdomain. Then, the intact and faulty regions are separated by an interface
$\Gamma := \partial \Omega_I \cap \partial \Omega_F$. 
We denote the unknowns $u_F,~u_I$ and $u_{\Gamma}$ with respect to the subdomains $\Omega_F,~\Omega_I$ and the interface $\Gamma$, respectively,
cf. Fig. \ref{fig:FaultTet}.

%
Moreover, we assume that a failure   occurs 
only after a complete multigrid cycle and is reported immediately.
The solution values 
in the faulty subdomain are set to zero as initial guess.
After the local re-initialization of the problem and a possible
recovery step, 
we continue with multigrid cycles in the solution process.
We focus on three categories of computational jobs in the solution process: A job in which
no fault occurs is called a {\em fault-free} job, a {\em do-nothing} job is a job in which
a failure occurs but no special recovery strategy is applied
(besides reinitializing the solution locally to zero),
and a {\em recovery} job stands for
a job where recovery strategies are applied after a fault happened.
We assume a process experiences a fault after $k_F$ cycles and count the number of iterations necessary 
to reach the stopping criteria (here: $10^{-15}$) in the case of a fault-free job $k_{\mbox{\small free}}$ and
a faulty job with possible recovery $k_{\mbox{\small faulty}}$. 
To quantify the extra iterations compared to a fault-free job,
we introduce a {\em relative Cycle Advantage} (CA) parameter $\kappa$ defined by
\begin{equation}\label{eq:CycleAdvantage}
 	\kappa :=  \frac{k_{\mbox{\small faulty}}-k_{\mbox{\small free}}}{k_F} .
\end{equation}
Intuitively, we expect that $\kappa \in [0,1]$.
The situation $\kappa = 0$ represents the case when the recovery algorithm reaches
the stopping criteria with the  same number of iterations
as in a fault-free job. For increasing $\kappa$,  more and more additional iterations are
required to reach the stopping criteria.
For $\kappa = 1$, $k_F $ additional  multigrid  iterations have to be carried out,
and this means that essentially all information that had been accumulated before the fault, has
been lost.

\begin{figure}[ht]
  	\pgfplotsset{tick label style ={font=\scriptsize}}
	\begin{tikzpicture}[scale= 0.6]
	\begin{semilogyaxis}
	[
		     width=0.65\textwidth,
		     height=0.5\textwidth,
			ylabel={Residual},
			legend pos=north east,
		xticklabels = {0, 5, 5, 10, 15, 20},
		xlabel={Iterations},
		xtick = {0, 5, 8, 13, 18, 23},
		ymajorgrids = true,
		xmin = 0, xmax = 23,
		ymin=1e-16,ymax=1,
		ytick = {1e0, 1e-4, 1e-8, 1e-12,1e-16},
		legend style={/tikz/every odd column/.style={yshift=2pt}, font=\scriptsize, text height = 0.7ex, legend cell align = left}
		]

		 \addplot[color=blue,mark=o] coordinates {
		(0.0, 1                    )
		(1.0, 0.026542107          )
		(2.0, 0.0013048188         )
		(3.0, 8.07622401223097E-005)
                (4.0, 0.000006283          )
                (5.0,6.23661729007494E-007 )
                };
\def \sh {3}
	\fill[text=FAULT, blue,fill opacity=0.25, ] (axis cs:5,1e-16) rectangle (axis cs: 8,1);        
	\node[rotate=90] at (axis cs:6.5,1e-8) {\Large \bf FAULT };             
                \addplot+[color=blue,mark=o , mark options = {fill=blue}] coordinates {
                (5 +\sh, 6.23621067526521E-007)
                (6 +\sh, 7.44060374166948E-008)
                (7 +\sh, 1.00529005867452E-008)
                (8 +\sh, 1.48514026177861E-009)
                (9+\sh, 2.34532575945481E-010)
                (10+\sh, 3.89048236714878E-011)
                (11+\sh, 6.6871464992498E-012 )
                (12+\sh, 1.17919108049753E-012)
                (13+\sh, 2.11872339214339E-013)
                (14+\sh, 3.87087946717195E-014)
                (15+\sh, 7.69193235556025E-015)
                (16+\sh, 3.22732207552463E-015)
                (17+\sh, 2.963831612675E-015  )
                (18+\sh, 2.95927752680608E-015)
                (19+\sh, 2.95576030870196E-015)
                (20+\sh, 2.961322799299E-015  )
                (21+\sh, 2.961322799299E-015  )
                 };
                 
        \addplot+[color=red, mark=square, mark options={solid}] coordinates { 
	      (5 +\sh, 0.1550719098		)
	      (6 +\sh, 0.0012707323             )
	      (7 +\sh, 5.97060988155311E-005    )
	      (8 +\sh, 0.000004814              )
	      (9 +\sh, 5.23915050033952E-007    )
	      (10+\sh, 6.7742433914928E-008     )
	      (11+\sh, 9.74131165805321E-009    )
	      (12+\sh, 1.50426742242806E-009    )
	      (13+\sh, 2.44211228261356E-010    )
	      (14+\sh, 4.1101438196582E-011     )
	      (15+\sh, 7.10315411107903E-012    )
	      (16+\sh, 1.25225976180504E-012    )
	      (17+\sh, 2.24183415808371E-013    )
	      (18+\sh, 4.07151541273436E-014    )
	      (19+\sh, 7.99140416292242E-015    )
	      (20+\sh, 3.25479703821773E-015    )
	      (21+\sh, 2.96864186587404E-015    )
               };

            \legend{ ,Fault-Free, Fault}
\end{semilogyaxis}
\end{tikzpicture}\hspace*{-3mm}
 \raisebox{8mm}{\includegraphics[width = 0.32\textwidth]{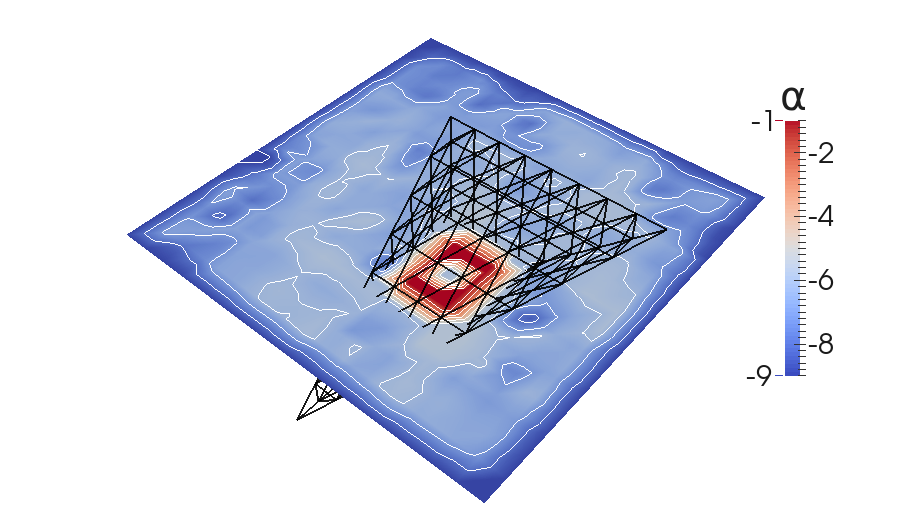}\hspace*{-4mm}
 \includegraphics[width = 0.32\textwidth]{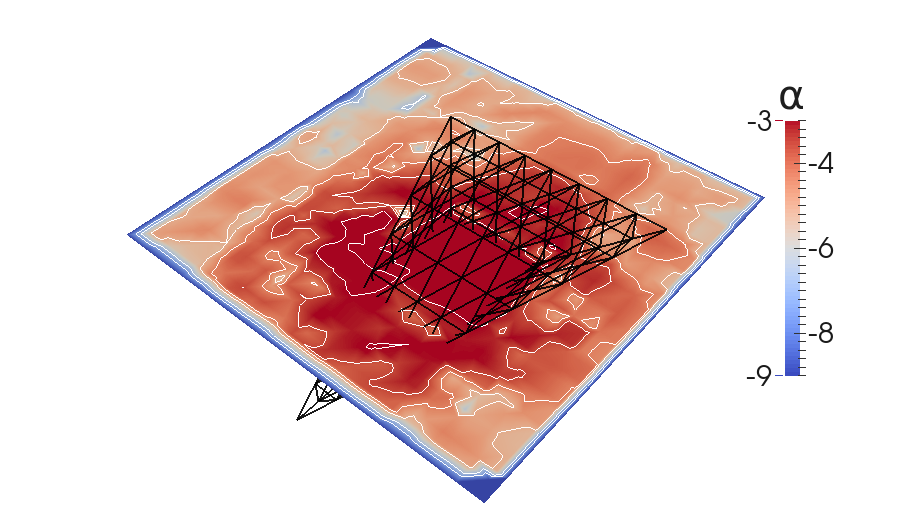}\hspace*{-18mm}}   
 \caption{Residual decay (left) and residual distribution after the failure (middle) and after one additional global V-cycle (right) restricted to a cross section through the domain $\Omega$, ($\alpha := \log_{10}(|Residual|)$).}
 \label{fig:ResPlot}
\end{figure}
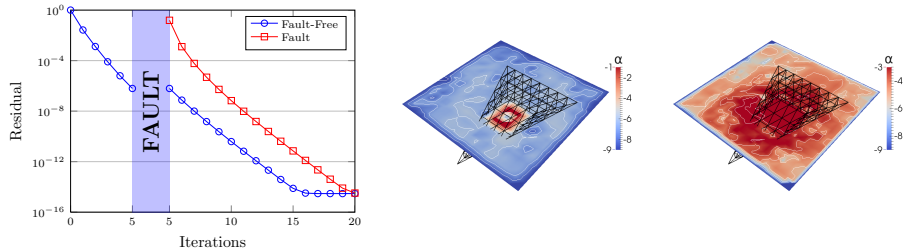

The following simple test setting illustrates the need for special recovery strategies. We consider
a fault scenario with 16 million unknowns and 
a loss of 0.3 million unknowns in case of a process crash in the unit cube $\Omega=(0,1)^3$.
The faulty subdomain is located similar to 
Fig. \ref{fig:FaultTet}. 
On the left of Fig.~\ref{fig:ResPlot}, we show the decay of the residual within a global parallel
multigrid scheme.
The fault occurs after 
$k_F = 5$ 
multigrid iterations and in the case of a do-nothing job, the residual
after the fault is highly increased resulting in 
four additional multigrid steps to obtain a
given tolerance compared to the fault-free job.
Further tests show that the number of additional steps is almost always
equal to $k_F -1$. Thus a do-nothing run results in a $\kappa $ close to one.
A favorable pre-asymptotically improved convergence rate after the fault often helps to save
one cycle, but besides from this, the extra cost incurred by the fault
is essentially the number of cycles that have been performed before the fault.
The situation becomes even worse 
when multiple faults occur.

The middle   
illustration in Fig.~\ref{fig:ResPlot} 
visualizes the residual  on a cross section
through the domain directly after the failure and re-initialization.
The residual distribution when one additional global multigrid cycle has been performed is given on the right.
Note that large local error components within $\Omega_F$ are 
dispersed globally by the multigrid cycle.
Although the smoother transports information only across a few neighboring mesh cells on each level,
their combined
contributions on all grid levels leads to a global pollution of the error.
Though the residual decreases globally, the residual
in the healthy domain $\Omega_I$ increases by this pollution effect.
A possible remedy is based on temporarily decoupling the domains
in order to avoid that the locally large residuals can pollute into the healthy
domain.
This is motivated by the asynchronous multilevel algorithms in
\cite{Ruede1993fully} and leads  to strategies that combine
domain decomposition \cite{Quarteroni,Bjoerstadt,ToselliWidlund} 
and multigrid techniques. 
They are essentially based on hierarchical and partitioned data structures
as will be discussed in the next section.

\section{Software requirement for algorithmic resilience}\label{sec:HHG}
In this section, we introduce 
a software architecture that is suitable to deal with faults and that supports 
numerical recovery procedures in the case of faults.
The main ingredient for the recovery algorithms is a 
hybrid data structure 
that allows to combine multigrid mesh hierarchies 
with tearing and interconnecting strategies from domain partitioning.
All our numerical results will be carried out within the HHG software
library, \cite{BergenHuelsemann04,GKSR14,GRSWW13}, 
but the essential techniques can be
adapted to other software concepts, since they essentially only exploit
the redundancy of data in ghost layers
as they are used also in other distributed memory parallel 
frameworks.

\subsection{Hybrid hierarchical grid data structures}

Often, the parallel communication of processes is 
organized in {\em ghost} layers (sometimes also called {\em halos})
which redundantly store copies of {\em master} data.
This is a convenient technique to accommodate data dependencies
across processor boundaries. 
Only the original {\em master} values can be written by the algorithm,
and thus, once a master value has been changed,
the associated ghost values must be updated to
hold consistent values.
We here propose a systematic construction of the ghost layer data structures
that is induced by the mesh geometry. 

\begin{figure}
 \centering
    \includegraphics[width = 0.9\textwidth]{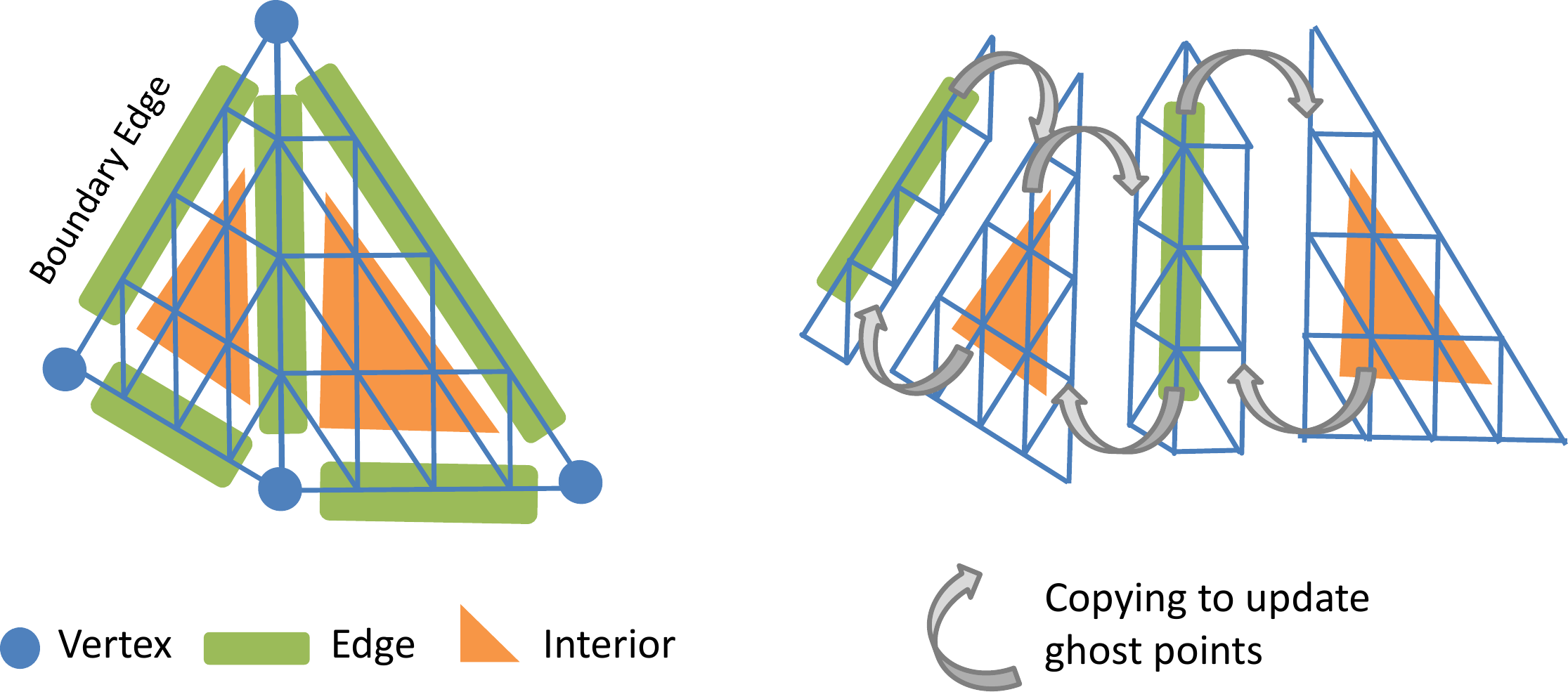} \\
\vspace{0.5cm}
    \includegraphics[width = 0.9\textwidth]{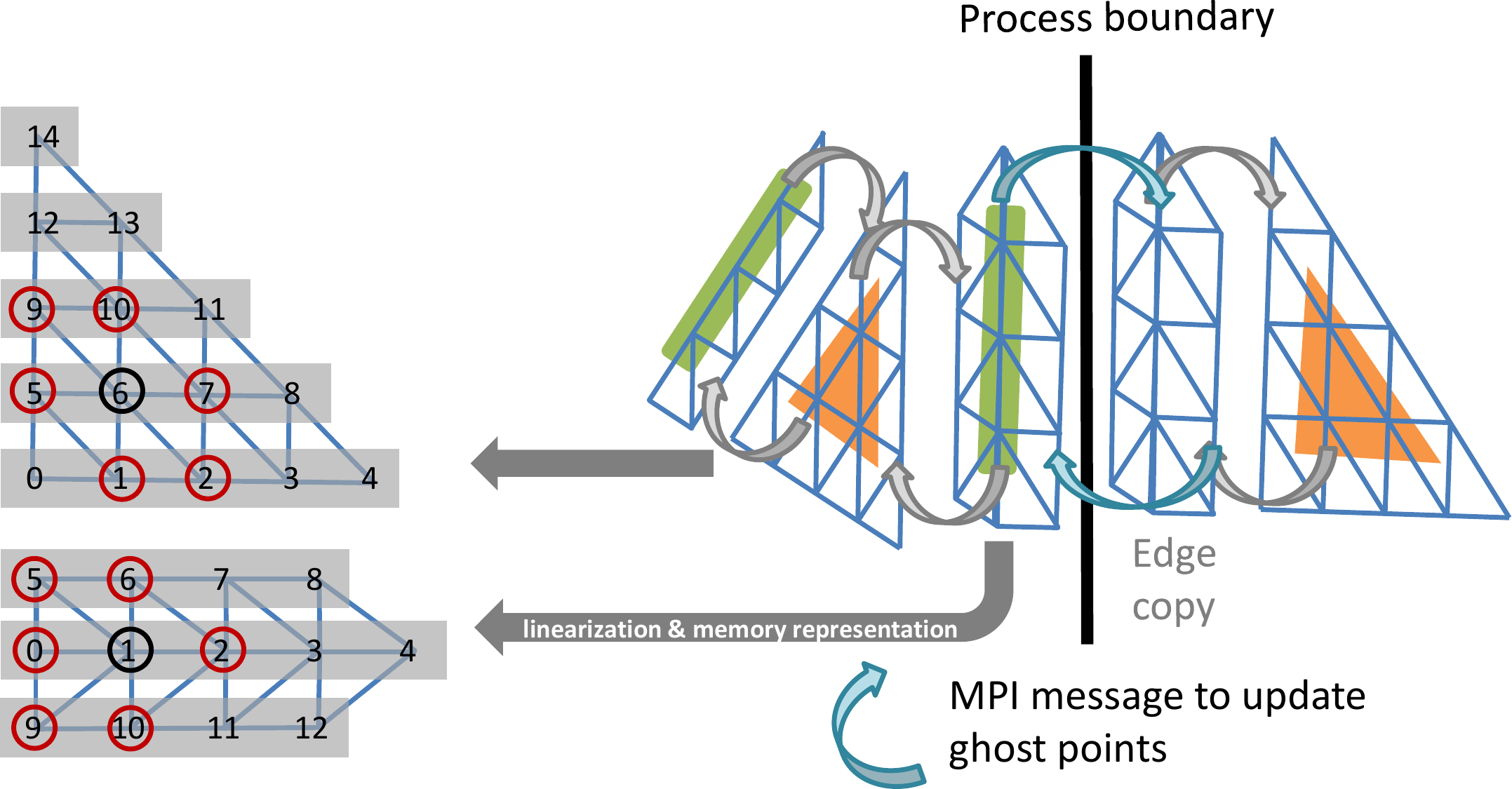}
   \caption{Top left: Two structured refined elements including grouping of the finite element nodes. Top right: Four containers for edge and interior nodes and updates of their corresponding ghost points. Bottom: Duplication of an edge container for a parallel example and synchronization via MPI. Memory representation of the data structures and one stencil for two different containers.}
 \label{fig:hhgstruct3}
\end{figure}

The tetrahedra of the unstructured coarsest mesh $\mathcal{T}_{-2}$
provide natural process boundaries for all refined meshes.
Note that the refined meshes will have nodes that are located on the vertices, edges, faces, and 
the interior of the initial elements in  $\mathcal{T}_{-2}$.
For two tetrahedra in 3D, this is illustrated in 
Fig.~\ref{fig:ghostlayer3d} below, but for the sake of simplicity we illustrate
the concept for the 2D case in Fig.~\ref{fig:hhgstruct3}.
The upper left of Fig.~\ref{fig:hhgstruct3}
illustrates how first all fine mesh nodes are classified.
In the upper right illustration, then the ghost layers are introduced.
This is achieved by separating the two connected triangular elements,
and identifying the common edge as a separate entity.
For the software architecture, this classification of nodes induces
a system of container data structures. In particular, there are 3D containers to hold
the nodes in the interior of each $T_i \in \mathcal{T}_{-2}$, then 2D containers for those nodes
that lie on each coarse mesh face $F_{i,j} =  T_i \cap T_j$, 1D containers for the nodes that lie
on the coarse mesh edges, and eventually 0D containers that hold trivially the nodes coinciding with the vertices of $\mathcal{T}_{-2}$. 
In a parallel setting, we will allocate  all these container objects onto the different processors. 

Conceptually, the next step is to introduce the ghost layers. The geometrical classification above
had distributed the {\em master} copies of each node into separate containers. As 
indicated in Fig.~\ref{fig:hhgstruct3} (top-right) (see also Fig.~\ref{fig:ghostlayer3d} for 3D),
these containers can be enriched by {\em ghost} nodes which are copies of
{\em master} nodes that are stored elsewhere. 
Thus all the fine grid nodes that rest on the boundary of a $T_i \in \mathcal{T}_{-2}$  
become ghost nodes in that $T_i$, similarly, in the face data structure, the nodes that lie on the edges become ghost nodes, and eventually the end points of edges become ghost nodes for the edge data container.
Furthermore, each of the face, edge, and node containers is enriched by one additional layer of 
ghost nodes that hold additional copies of master values.
These extra ghost layers are essential for e.g.\ efficiently implementing a Gauss-Seidel iteration
for the master nodes of the corresponding face container.

For the parallel implementation, we eventually introduce additional
copies of the face, edge, and vertex 
containers so that they can be
uniquely allocated on the processors of a distributed memory 
system.
In the HHG implementation, the communication between nodes is thus split on two subtasks.
One consist in updating the ghost node values in the containers that are local to a processor by simple
local copy operations.
The second is the actual synchronization of the interface data structures (face, edge, node)
by message passing in the distributed memory system.

We remark that this construction obviously leads to a data
redundancy and extra cost in data storage, but that for significantly refined meshes the
additional memory cost is of lower order complexity.
As we will see below, this redundancy enables efficient implementations of the parallel multilevel 
solver algorithms, including a systematic recovery after a fault, i.e., the numerical recovery of the data
in these structures.

Note also, that these containers can be implemented using  
simple array data structures and can be accessed with integer indices.
This can be systematically exploited to design a highly efficient implementation that 
avoids indirect addressing and that permits efficient vectorizable looping through the data,
as indicated in Fig.\ \ref{fig:hhgstruct3} (lower left). 
Further note, that the interface data structures can also be used to implement Dirichlet or Neumann boundary conditions for those tetrahedra that lie at the domain boundary, see also Subsec. \ref{subsec:algebraic} for the resulting linear algebraic structure.
We point out  that in case of a fault, the master nodes  of a tetrahedral container may be lost
and no copies of the master values will exist anywhere.
However, clearly for all of the interface structures, the above construction will
provide ghost copies, from which the data can be restored.

In the following, we will design suitable recovery algorithms for this task.
As a first step, we will study the resulting linear algebra structures.

\subsection{Equivalent algebraic formulations} \label{subsec:algebraic}
The data structure discussed in the previous subsection
allows different equivalent algebraic formulations of \eqref{eq:PoissonLin}
that play an important role in the following sections.
\begin{figure}[ht]
\centerline{
\hspace*{-1.5cm}
\includegraphics[width=0.55\textwidth]{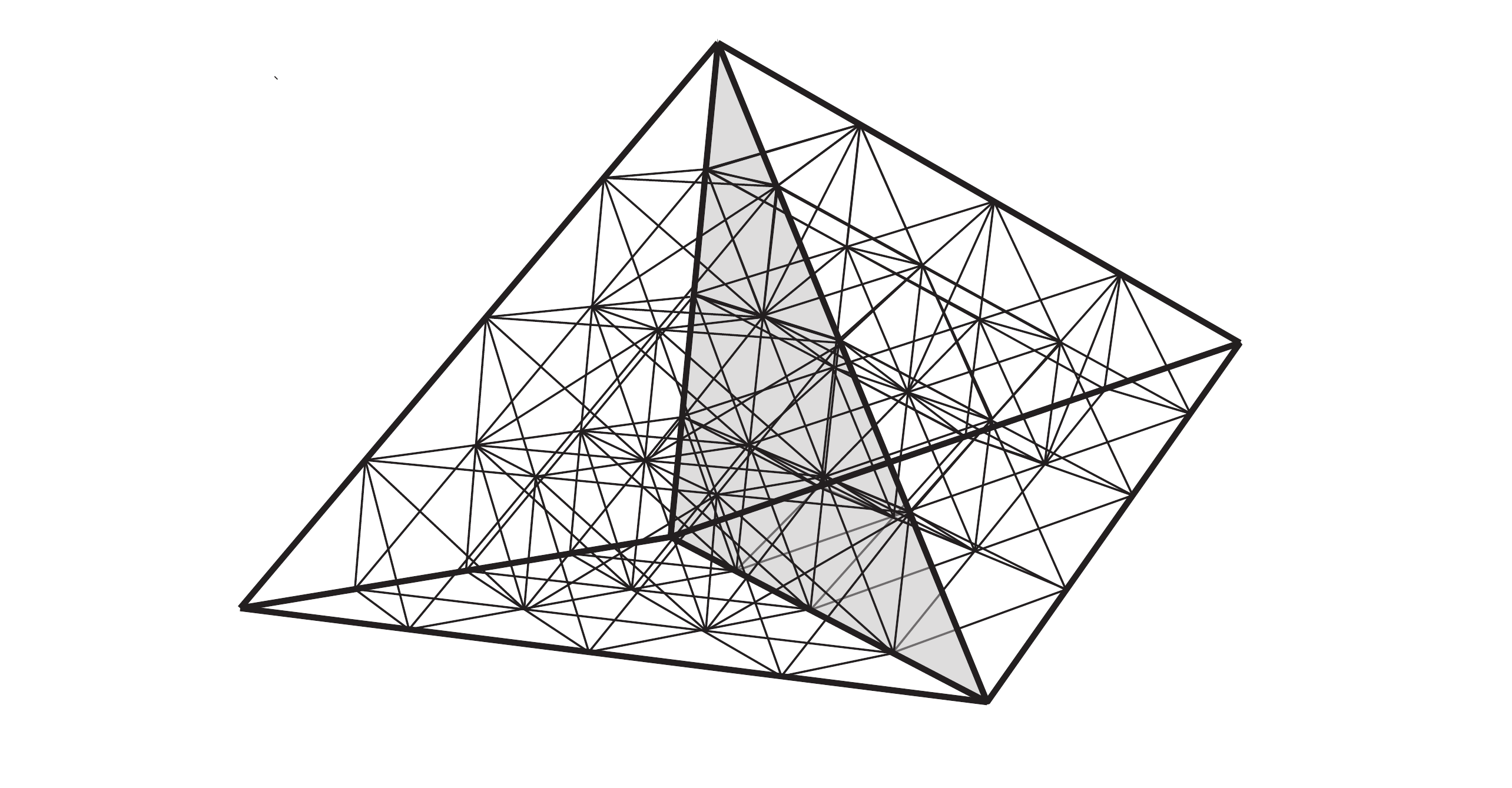}
\hspace*{-0.6cm}
\includegraphics[width=0.55\textwidth]{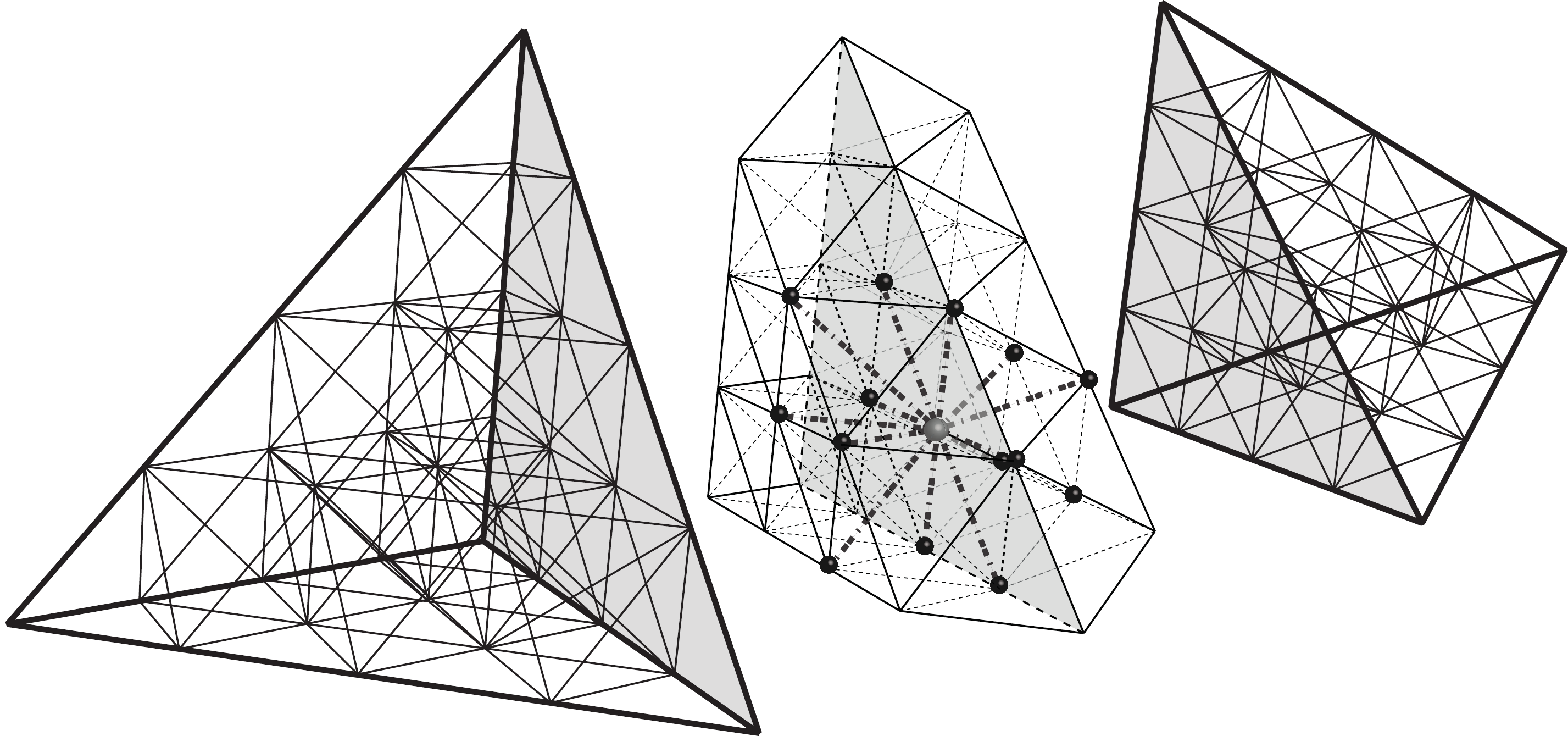}}
\caption{Two neighboring master elements (left) and ghost-layer
  structure of the common master face (right)}
\label{fig:ghostlayer3d}
\end{figure}
Fig. \ref{fig:ghostlayer3d} illustrates the ghost-layer structure of a
coarse mesh face with the master and ghost nodes of a refined
mesh. For the interface container structures (faces, edges, vertices), a recovery is directly possible, since we can assume that redundant copies exist in the system.
However, for the volume elements (tetrahedra),
their inner nodes are not available, and thus they must be re-constructed numerically.
The ghost layer structure and the associated data redundancy
can be represented algebraically by rewriting \eqref{eq:PoissonLin}
equivalently as
\begin{equation}\label{eq:algSystemGhost}
 \begin{pmatrix}
  A_{II}  & A_{I\Gamma_I} & {\bf 0} & {\bf 0} & {\bf 0} \\
  {\bf 0} & \text{\bf Id} & -\text{\bf Id} & {\bf 0} & {\bf 0}\\
  A_{\Gamma I} & {\bf 0} & A_{\Gamma \Gamma} & {\bf 0} & A_{\Gamma F} \\
  {\bf 0} & {\bf 0} & -\text{\bf Id} & \text{\bf Id} & {\bf 0} \\
   {\bf 0} & {\bf 0} & {\bf 0} & A_{F \Gamma_F} & A_{F F} \\
 \end{pmatrix}
 \begin{pmatrix}
  \ul u_I\\
  \ul u_{\Gamma_I}\\
  \ul u_{\Gamma}\\
  \ul u_{\Gamma_F}\\
  \ul u_{F}\\
 \end{pmatrix}
 = 
  \begin{pmatrix}
  \ul f_I\\
   {\bf 0}\\
   {\bf 0}\\
   {\bf 0}\\
  \ul f_{F}\\
 \end{pmatrix} .
\end{equation}
The sub-matrices are associated with the block unknowns, in more general settings,
they also depend on the basis functions and the PDE.
Because of the locality of the support of the basis function, we can identify $A_{\Gamma I}$ with
$A_{I \Gamma_I}^\top$ and $ A_{\Gamma F}$ with $A_{F \Gamma_F}^\top$.
Row 2 and row 4 in \eqref{eq:algSystemGhost} guarantee the consistency of the redundant data
at the interface between the master elements. Row 3 reflects,
the ghost layer structures associated with the master faces.
We recall that by our assumptions 
the data $\ul u_F$ and $\ul u_{\Gamma_F}$ are lost but
$\ul u_I$ and  $\ul u_{\Gamma_I} $  are still available.
Although it is a priori not known what the intact domain will
be, the data structure allows always such a presentation.
If a processor fails which is associated with master faces, then the
recovery for $\ul u_\Gamma $ can be trivially performed using 
the redundant information of row 2.
Thus, we focus in the following only on the recovery of $\ul u_F$ and
$\ul u_{\Gamma_F}$.

\begin{figure}[ht]
\centerline{
\includegraphics[width=0.25\textwidth]{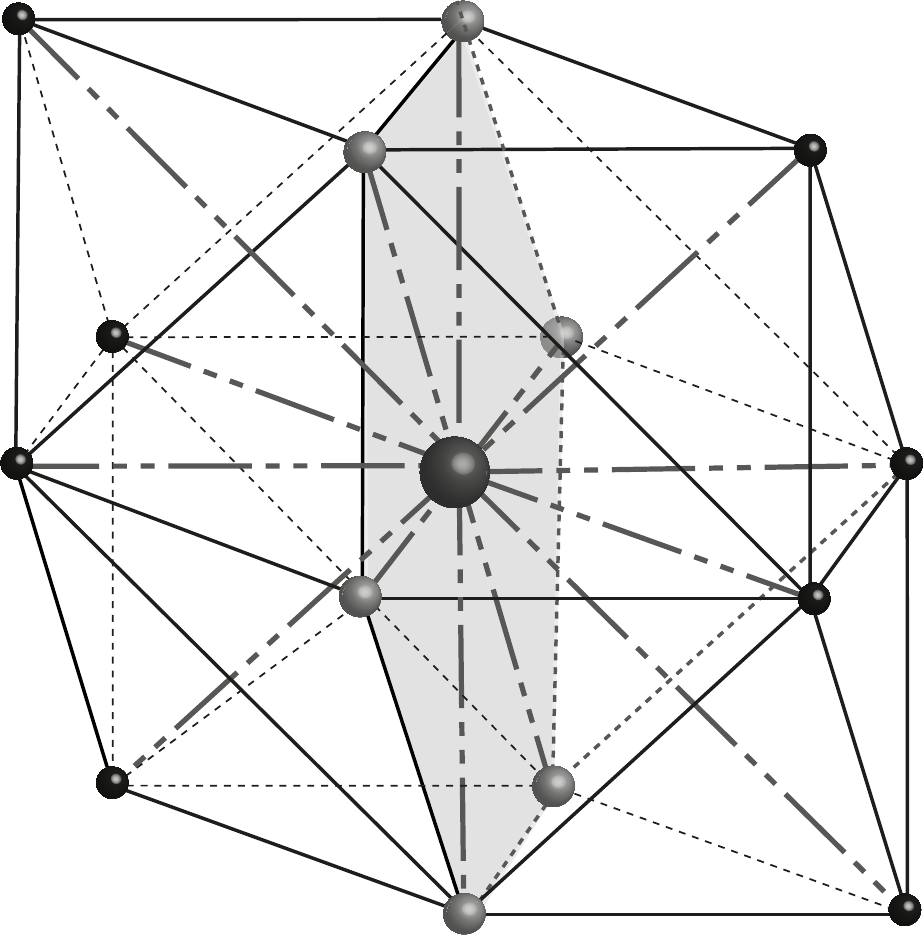}
\hspace*{2cm}
\includegraphics[width=0.35\textwidth]{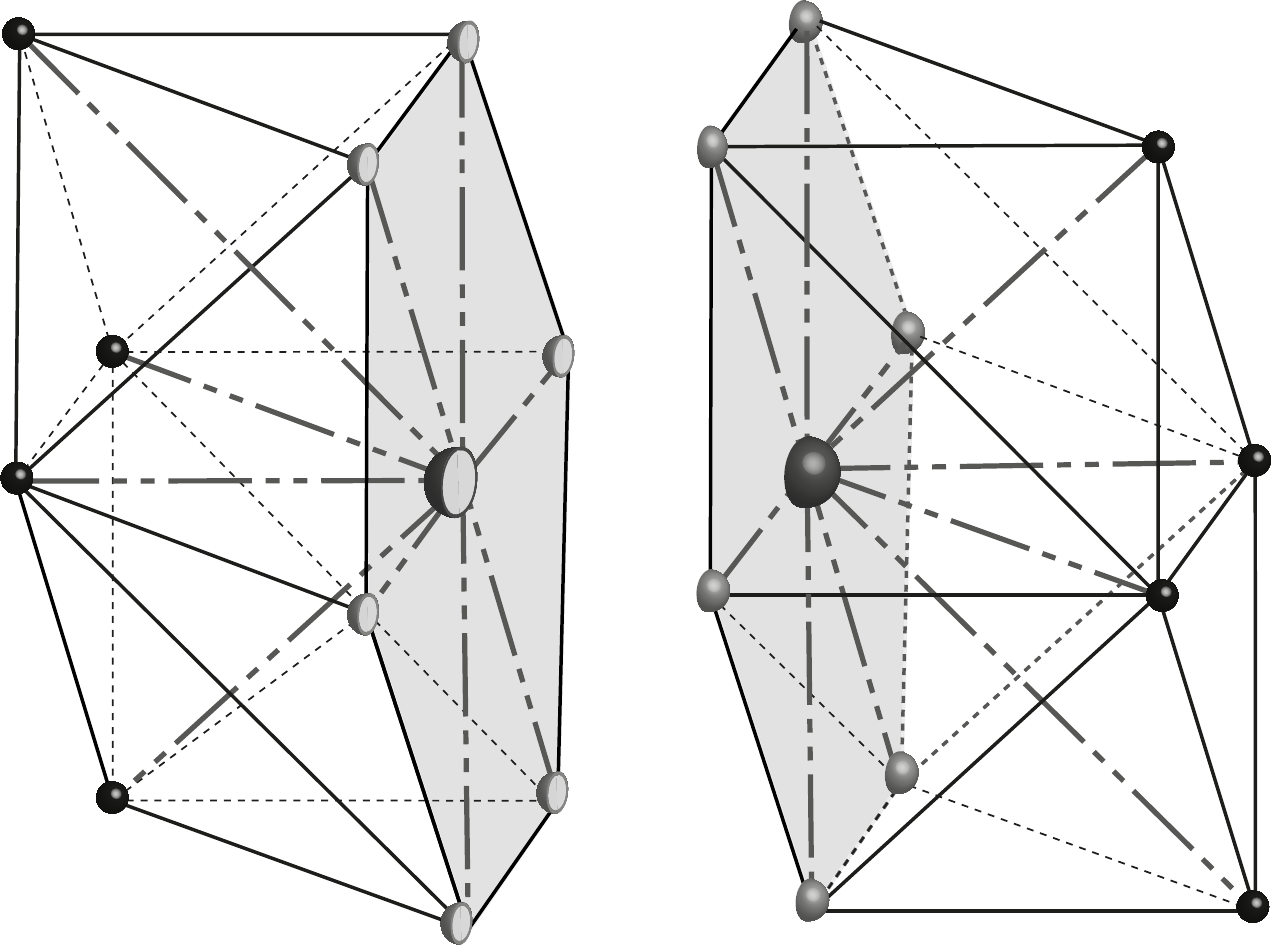}}
\caption{Stencil (left) and sub-stencil (right) structure for a vertex on a master face}
\label{fig:halfstencil}
\end{figure}

The 
ghost layer structure does not only permit access to 
$A_{\Gamma \Gamma} $ but also to decompose it into 
$A_{\Gamma_F \Gamma_F} + A_{\Gamma_I \Gamma_I} $.
More precisely for each vertex on a master face, the two sub-stencils
associated with $\Omega_I$ and $\Omega_F$
are available, see also Fig. \ref{fig:halfstencil}.
Having the sub-stencils at hand, we can rewrite 
\eqref{eq:PoissonLin} equivalently 
in terms of the additional flux unknown $\ul \lambda_{\Gamma_I}$ as
\begin{equation}\label{eq:algSystemGhostN}
\begin{pmatrix}
  A_{II}  & A_{I\Gamma_I} & {\bf 0} & {\bf 0} & {\bf 0} & {\bf 0} \\
  A_{\Gamma_I I}  & A_{ \Gamma_I \Gamma_I} & {\text{\bf Id}} & {\bf 0} & {\bf 0} & {\bf 0} \\
  {\bf 0} & {\text{\bf Id}} & {\bf 0}& -{\text{\bf Id}} & {\bf 0} & {\bf 0}\\
  A_{\Gamma I} & {\bf 0} & {\bf 0} & A_{\Gamma \Gamma} & {\bf 0} & A_{\Gamma F} \\
  {\bf 0} & {\bf 0} & {\bf 0} & -{\text{\bf Id}} & {\text{\bf Id}} & {\bf 0} \\
   {\bf 0} & {\bf 0} & {\bf 0} & {\bf 0} & A_{F \Gamma_F} & A_{F F} \\
 \end{pmatrix}
 \begin{pmatrix}
  \ul u_I\\
  \ul u_{\Gamma_I}\\
  \ul \lambda_{\Gamma_I}\\
  \ul u_{\Gamma}\\
  \ul u_{\Gamma_F}\\
  \ul u_{F}\\
 \end{pmatrix}
 = 
  \begin{pmatrix}
  \ul f_I\\
  {\bf 0}\\
   {\bf 0}\\
   {\bf 0}\\
   {\bf 0}\\
  \ul f_{F}\\
\end{pmatrix} .
\end{equation}
Here $-\ul \lambda_{\Gamma_I}$ stands for the discrete flux out of
$\Omega_I$ and into $\Omega_F$.
It reflects a Neumann boundary condition for the intact, healthy domain.
In \eqref{eq:algSystemGhost} and \eqref{eq:algSystemGhostN}, we thus find
the typical algebraic structure of Dirichlet and Neumann subproblems
associated with the faulty and the intact subdomain, respectively.
This observation motivates the design of our  parallel recovery  algorithms.
%

\section{Local recovery strategy}\label{sec:localrecovery}

So far we have only considered 
local recovery strategies in the faulty subdomain.
During 
the recovery, the processes in the intact subdomain 
are halted until the recovery in the faulty subdomain has finished.
However, for the overall efficiency we cannot neglect the time spent 
in the recovery process.
Thus, we extend our approach into two directions. Firstly, we
introduce a superman strategy to speed up the local recovery process
itself,
and secondly we propose  a global recovery such that on the faulty and
intact subdomains parallel but asynchronous processes take place.

\subsection{The local superman}
In parallel geometric multigrid, load balancing can be often carried out
with respect to the degrees of freedom. Here the situation is quite
different.
The 
multigrid iteration in the faulty subdomain  
must deal with a much higher residual compared to the
residual on the healthy 
domain. To compensate for this, we use a local {\em superman} 
processor,  i.e.\ we {\em over-balance} the compute power in the faulty subdomain.
Technically, the additional compute resources  
for realizing the superman 
can be 
provided by additional parallelization.
We propose here, that e.g.\ a full (shared memory) compute node is
assigned to perform the local
recovery for a domain that was previously handled by a single core.
This can be accomplished by a tailored OpenMP parallelization for the recovery process.
Alternatively, a further domain partitioning of $\Omega_F$ can be
used together with an additional local distributed memory parallelization.
Alternatively, 
one may think to exploit accelerators, such as GPUs or a Xeon Phi. 

On the other hand
in practice, $\Omega_F$ is much smaller than $\Omega_I$ and 
will typically fit in a single processor memory. 
Thus a suitably designed recovery
process 
will automatically benefit from locality.
In this case, the local process 
will be automatically faster than a global process of the same type involving full message
passing communication over long distances. 
Moreover, 
in a standard multigrid implementation, the coarsest grid
for $\Omega_F$ contains only one inner node, and thus the coarse grid problem
within the local multigrid cycle is trivial to solve and does not
contribute to the time per multigrid cycle.
This is in contrast to large scale
parallel computations,
when the cost of the coarse grid solver can often not be neglected,
since it may 
contribute $50\%$ or more of the total time to solution. 

Since we here assume that local and global computation are performed simultaneously, 
the time for the recovery is  
presented as the maximum of the time spent in $\Omega_I$ 
and in $\Omega_F$. 
By introducing  $\eta_{\super}$ as  the acceleration ratio of the cycles   
that can be performed simultaneously in $\Omega_F$,
parallel to one iteration  
in $\Omega_I$, we obtain a quantitative measure for the expected
speed up.
This speedup can  
be caused 
by an increase of the compute power and/or an
decrease of the communication overhead and a complexity decrease in the
coarse grid solver.
If $\eta_{\super} = 1$, there is no speedup, i.e., one global cycle can be performed as fast as a
local one. For the ideal but practically not relevant situation $\eta_{\super} = \infty$, 
the local recovery of Sec. \ref{sec:localrecovery} does not
contribute to the global run time.

\subsection{Dirichlet-Dirichlet recovery strategy}\label{sec:DDR}
In the Dirichlet-Dirichlet recovery, 
we freeze the values $\ul u_{\Gamma_I}$ at the interface $\Gamma_I$. This allows to compute
the two subproblems in $\Omega_F$ and $\Omega_I$ independently and consequently no communication between $\Omega_F$ and $\Omega_I$
is necessary.
Thus, it is guaranteed that no defect data is  
polluted into the healthy subdomain during 
the recovery. On both subdomains, we  
iterate now on decoupled Dirichlet problems with boundary data on
$\Gamma$ given by $\ul u_\Gamma$.
Obviously, at some point we have to 
reconnect the two subdomains again.  
The algorithm for the Dirichlet-Dirichlet recovery is presented in
Alg.\ \ref{alg:DDAlg}

\begin{algorithm}
\caption{Dirichlet-Dirichlet recovery (DD) algorithm}\label{alg:DDAlg}
\begin{algorithmic}[1]
    \State Solve \eqref{eq:PoissonLin} by multigrid cycles.
    \If{Fault has occurred}
    \State \textbf{STOP} solving \eqref{eq:PoissonLin}.
    \State Recover boundary data $\ul u_{\Gamma_{F}}$ from line 4 in  \eqref{eq:algSystemGhost}
    \State Initialize $\ul u_F $ with zero
    \State \textbf{In parallel do:}
    \State ~a) Use $n_F$ MG cycles accelerated by $\eta_{\super}$
    	to approximate line 5  in \eqref{eq:algSystemGhost}:
\State  $\qquad A_{FF} \ul u_F = \ul f_F - A_{F\Gamma_F} \ul u_{\Gamma_F}$
    \State ~b)  Use $n_I$ MG cycles to approximate line 1 in  \eqref{eq:algSystemGhost}
\State  $\qquad A_{II} \ul u_I = \ul f_I - A_{I\Gamma_I} \ul u_{\Gamma_I}$
    \State \textbf{RETURN} to line 1 with new values $\ul
    u_I$ in $\Omega_I$ and $\ul u_F$ in $\Omega_{F}$.
    \EndIf
  \end{algorithmic}
\end{algorithm}

\subsection{Dirichlet-Neumann recovery strategy}\label{sec:DNR}
In the Dirichlet-Neumann recovery strategy, we do not freeze the interface values  
but treat them as Neumann boundary data in the healthy 
subdomain.
By doing so, we use a  one-directional coupling of the faulty subdomain 
$\Omega_F$ and the healthy 
subdomain $\Omega_I$. 
On  $\Omega_I$ 
we approximate a Neumann boundary problem with static data,
whereas on $\Omega_F$, we approximate a  Dirichlet problem 
with dynamic boundary data. 
After each multigrid cycle on $\Omega_I$,  the newly computed interface values
$\ul u_{\Gamma_I}$ are asynchronously communicated via $\ul u_{\Gamma} $ onto 
$\ul u_{\Gamma_{F}}$. 
Hence, we only avoid communication from the  faulty to the healthy domain 
but still keep communicating from the intact 
to the faulty subdomain. 
As in the case of  the Dirichlet-Dirichlet recovery strategy, it is necessary to fully interconnect both subdomains  after a couple of cycles.
The algorithm is presented in Alg.\ \ref{alg:DNAlg}.

\begin{algorithm}
\caption{Dirichlet-Neumann recovery (DN) algorithm}\label{alg:DNAlg}
\begin{algorithmic}[1]
    \State Solve \eqref{eq:PoissonLin} by multigrid cycles.
    \If{Fault has occurred}
    \State \textbf{STOP} solving \eqref{eq:PoissonLin}.
    \State Compute the Neumann condition from line 2 in    \eqref{eq:algSystemGhostN}: 
    \State $\qquad	     \ul \lambda_{\Gamma_I} = - A_{\Gamma_I I} \ul u_I -
             A_{\Gamma_I \Gamma_I}  \ul u_{\Gamma_I} $
    \State  Initialize $\ul u_F$ with zero.
    \State \textbf{In parallel do:}
     \State ~a) Perform $n_F$ cycles with  acceleration $\eta_{\super}$:
 \State $\qquad$ Recover/update boundary values $\ul u_{\Gamma_{F}}$ from line 5  in
 \eqref{eq:algSystemGhostN}.
\State
$\qquad $ Use 1 MG cycle to approximate line 6  in  \eqref{eq:algSystemGhostN}:
\State  $\qquad A_{FF} \ul u_F = \ul f_F - A_{F\Gamma_F} \ul
u_{\Gamma_F}$
    \State ~b)  Perform $n_I$ cycles: 
\State $\qquad $  Use 1 MG cycle to approximate lines 1+2  in  \eqref{eq:algSystemGhostN}
\State $\qquad 
 \begin{pmatrix}
      A_{II} & A_{I \Gamma_I}\\
      A_{\Gamma_I I} & A_{\Gamma_I \Gamma_I} \\
     \end{pmatrix}
     \begin{pmatrix}
      \ul u_I\\
      \ul u_{\Gamma_I}
     \end{pmatrix}
     =
     \begin{pmatrix}
      \ul f_I \\
      \ul \lambda_{\Gamma_I}\\
     \end{pmatrix}
   $
\State $\qquad$ Update boundary values $\ul u_{\Gamma}$ from line 3  in
 \eqref{eq:algSystemGhostN}.
   \State \textbf{RETURN} to line 1 with values $\ul
    u_I$ in $\Omega_I$, $\ul u_\Gamma $ on $\Gamma$ and $\ul u_F$ in $\Omega_{F}$.
    \EndIf
  \end{algorithmic}
\end{algorithm}
\subsection{Comparison of the recovery strategies}
\label{sec:comprecstrats}
In this subsection, we compare the Dirichlet-Dirichlet (DD) and Dirichlet-Neumann (DN) recovery with the local recovery strategy (LR),
where calculations are only performed in the faulty subdomain while the healthy domain stays idle.

\begin{figure}
 \centering
    \includegraphics[width = 0.75\textwidth]{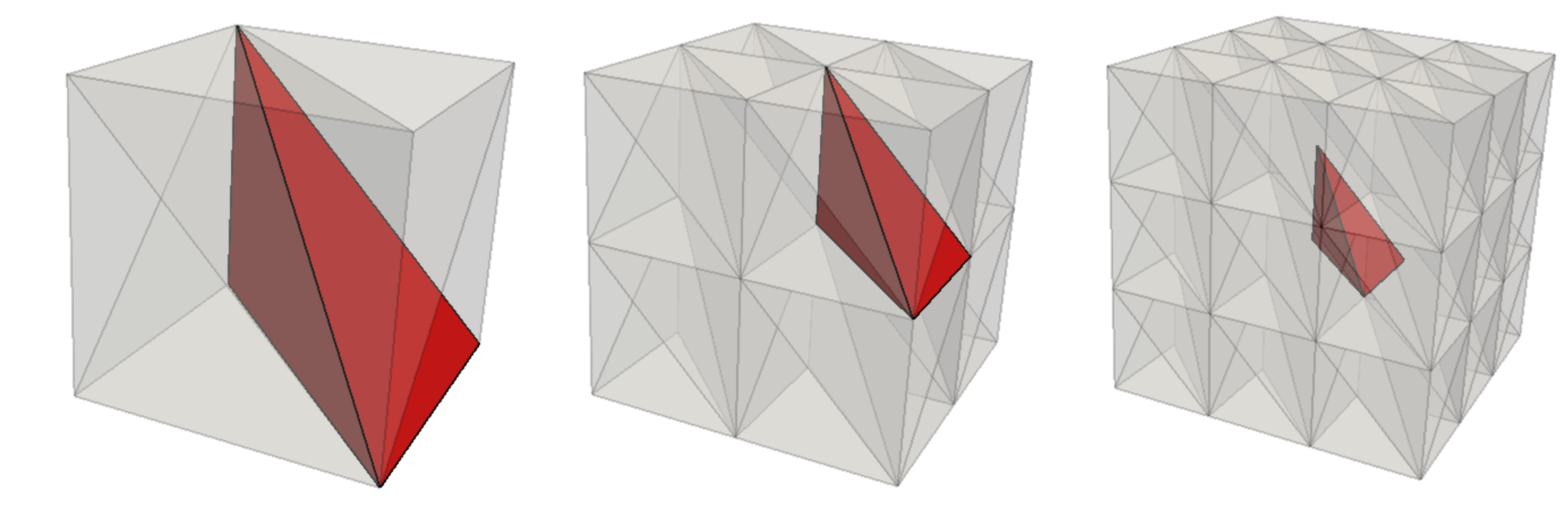}
   \caption{From left to right: Scenario (I), Scenario (II) and Scenario (III)}
 \label{fig:StudyFault}
\end{figure}

From the left to the right in Fig. \ref{fig:StudyFault}, the three different
fault geometries and macro-meshes  are shown which is uniformly
refined seven times for the following computations.
Scenario~(I)  
has  2 millions unknowns and the fault (red marked 
tetrahedron) is located with two faces at the boundary of the computational
domain and affects about 16.7\% of the total unknowns. Scenario~(II)
has 16 millions unknowns, a corruption by the fault of 2.0\% (like in Subsec. \ref{sec:AlgoPerf}),
and the faulty domain has one edge  
coinciding with the boundary.
Scenario (III) has  56 millions unknowns, 
the faulty domain is floating in the center of the $\Omega$ and 
affects only 0.6\% of the unknowns. 
In Tab. \ref{tab:ComparisonFault5} and Tab. \ref{tab:ComparisonFault11}, we  
present the values 
for the cycle advantage $\kappa$ with respect to the three different fault scenarios. 
We compare 
these strategies for two different superman speed up factors
$\eta_{\super} = 2$ and $\eta_{\super} = 5$. In addition, we also  illustrate the influence  of  $n_I$
in the recovery strategy before interconnecting the faulty and intact domain.
Please note that the choice $n_I=0$ reflects a do-nothing job.
We always fix $ n_F =
\lceil n_I * \eta_{\super} \rceil$ and run the numerical tests for $n_I \in
\{0,1,2,3,4\}$, i.e., in the case $n_I=3$ and $\eta_{\super} =2$, we have 
$n_F = 6$. To evaluate $\kappa$, we set $k_{\text{faulty}} $  to the sum of the required global multigrid
cycles and $n_I$. We point out that in Sec. \ref{sec:localrecovery}, we had $\eta_{\super} =1$, i.e., $ n_I = n_F$, and we did not add $n_I$ to the global multigrid cycle to get $k_{\text{faulty}} $. From the results of Sec. \ref{sec:localrecovery}, it is now obvious that the local recovery strategy without $\eta_{\super} > 1$ is of no interest.
\begin{table}[ht]
  \footnotesize\centering
  \caption[Tab.]{Comparison of the Cycle Advantage $\kappa$ for 
                 the local, Dirichlet-Dirichlet and Dirichlet-Neumann recovery strategy for a fault after 5 iterations.}
  \label{tab:ComparisonFault5}
    \vspace{1em} 
\begin{tabular}{c|ccc|ccc|ccc}
  \toprule
  \multicolumn{10}{c}{{\bf Superman speedup} $\eta_{\super}=2$}\\
  \hline
	 & \multicolumn{3}{c|}{Scenario (I)}	& \multicolumn{3}{c|}{Scenario (II)}	& \multicolumn{3}{c}{Scenario (III)}\\
  \hline
  $n_I$	  	&  LR 		&	DD	  &       DN	&   	 LR 		&	DD	  &    DN            	&  LR 		&	DD	  &    DN     \\      
\hline
 0	&	0.60	&	0.60	&	0.60	&    0.80	&	0.80	&	0.80	&	0.80	&	0.80	&	0.80	\\
									   													
1	&	0.20	&	0.00	&	0.00	&    0.40	&	0.40	&	0.40	&	0.20	&	0.20	&	0.00	\\
									   													
2	&	0.40	&	0.20	&	0.00	&    0.40	&	0.40	&	0.20	&	0.40	&	0.00	&	0.00	\\
									   													
3	&	0.60	&	0.40	&	0.20	&    0.60	&	0.60	&	0.40	&	0.60	&	0.20	&	0.00	\\
									   													
4	&	0.80	&	0.60	&	0.40	&    0.80	&	0.80	&	0.60	&	0.80	&	0.40	&	0.20	\\
\bottomrule
\end{tabular}
\vspace*{5pt}
%
\begin{tabular}{c|ccc|ccc|ccc}
  \toprule
   \multicolumn{10}{c}{{\bf Superman speedup}  $\eta_{\super}=5$}\\
   \hline
  $n_I$	  	&  LR 		&	DD	  &      DN 	&    	 LR		&	DD	  &    DN           	  	&  LR 		&	DD	  &    DN    \\       
\hline
  0	&   0.60	&	0.60	&	0.60	&	0.80	&	0.80	&	0.80	&		0.80	&	0.80	&	0.80\\
	                                                 		                            		
1&	    0.20	&	0.00	&	0.00	&	0.20	&	0.20	&	0.20	&		0.20	&	0.00	&	0.00\\
	                                                 		                            		
2&	    0.40	&	0.20	&	0.00	&	0.40	&	0.40	&	0.20	&		0.40	&	0.00	&	0.00\\
	                                                 		                            		
3&	    0.60	&	0.40	&	0.20	&	0.60	&	0.60	&	0.40	&		0.60	&	0.20	&	0.00\\
	                                                 			                            		
4&	    0.80	&	0.60	&	0.40	&	0.80	&	0.80	&	0.60	&		0.80	&	0.40	&	0.20\\
\bottomrule
\end{tabular}
\end{table}
In Tab. \ref{tab:ComparisonFault5}, the results are given for a fault after 
5 global multigrid iterations.
All recovery strategies can drastically reduce the delay
compared to a do-nothing strategy 
since they can attain a considerably smaller cycle advantage $\kappa$. 
For Scenario (I) and Scenario (III), we can even achieve $\kappa=0$.
The Scenario (III) is more relevant for applications on mid-size clusters,
whereas Scenario (I) is more relevant for small desktop systems.
Relevant cases for peta-scale systems will be considered in the next section.
We observe a clear advantage in using a global recovery strategy instead of a local recovery strategy.
In the case of the (LR) strategy, most of the machine stays idle for $n_I$ multigrid cycles.
Only in the case of the two global recovery strategies, we fully exploit the compute
power of the total system.
The Dirichlet-Neumann recovery strategy is the most powerful and robust
one with respect to the choice of $n_I$.
In all cases, $n_I$ plays a crucial role in the overall performance. If $n_I$ is too large, we are over-solving the subdomain problems,
if $n_I$ is too small, then the approximation on the faulty subdomain is too inaccurate and the global performance
suffers from the pollution of the
local defect into the global subdomain 
%
%
\begin{table}[ht]
  \footnotesize\centering
  \caption[Tab.]{Comparison of the Cycle Advantage $\kappa$ for 
                 the local, Dirichlet-Dirichlet and Dirichlet-Neumann recovery strategy for a fault after 11 iterations.}
  \label{tab:ComparisonFault11}
    \vspace{1em} 
  
     \begin{tabular}{c|ccc|ccc|ccc}
  \toprule
   \multicolumn{10}{c}{{\bf Superman speedup}  $\eta_{\super}=2$}\\
   \hline		 & \multicolumn{3}{c|}{Scenario (I)}	& \multicolumn{3}{c|}{Scenario (II)}	& \multicolumn{3}{c}{Scenario (III)}\\
  $n_I$	  	&  LR 		&	DD	  &      DN 	&    	 LR 		&	DD	  &    DN           	  	&  LR 		&	DD	  &    DN    \\
\hline

0	&	0.82	&	0.82	&	0.82	&	0.91	&	0.91	&	0.91	&	0.91	&	0.91	&	0.91 \\
																					
1	&	0.64	&	0.64	&	0.63	&	0.82	&	0.82	&	0.82	&	0.64	&	0.64	&	0.64  \\
																					
2	&	0.55	&	0.55	&	0.55	&	0.55	&	0.55	&	0.55	&	0.45	&	0.45	&	0.45 \\
																					
3	&	0.36	&	0.36	&	0.36	&	0.45	&	0.45	&	0.45	&	0.36	&	0.36	&	0.36 \\
																					
4	&	0.36	&	0.27	&	0.27	&	0.36	&	0.36	&	0.36	&	0.36	&	0.27	&	0.27  \\

5	&	0.45	&	0.36	&	0.27	&	0.45	&	0.45	&	0.36	&	0.45	&	0.27	&	0.27  \\

6	&	0.64	&	0.45	&	0.36	&	0.55	&	0.55	&	0.45	&	0.45	&	0.36	&	0.36  \\

\bottomrule
  \end{tabular}
  
  \vspace*{5pt}

   \begin{tabular}{c|ccc|ccc|ccc}
   \toprule
    \multicolumn{10}{c}{{\bf Superman speedup} $\eta_{\super}=5$}\\
      \hline
   $n_I$	  	&  LR 		&	DD	  &      DN 	&    	 LR 		&	DD	  &    DN           	  	&  LR 		&	DD	  &    DN    \\ 
\hline 

0	&	0.82	&	0.82	&	0.82	&	0.91	&	0.91	&	0.91	&	0.91	&	0.91	&	0.91\\
			
1	&	0.36	&	0.36	&	0.36	&	0.36	&	0.36	&	0.36	&	0.27	&	0.27	&	0.27\\
			
2	&	0.18	&	0.09	&	0.00	&	0.18	&	0.18	&	0.09	&	0.18	&	0.00	&	0.00\\
			
3	&	0.27	&	0.18	&	0.09	&	0.27	&	0.27	&	0.18	&	0.27	&	0.09	&	0.09\\
			
4	&	0.36	&	0.27	&	0.18	&	0.36	&	0.36	&	0.27	&	0.36	&	0.18	&	0.18\\

\bottomrule
\end{tabular}
\end{table}
%
%

To obtain a better feeling for the optimal $n_F$ depending on $k_F$, we consider in
Tab.~\ref{tab:ComparisonFault11} the situation of a fault after 11 multigrid cycles.
In contrast to Tab.~\ref{tab:ComparisonFault5}, we 
must select $n_I$ larger to get the  
best cycle advantage $\kappa$. 
This results from the fact that now 
the local subproblem solver on the faulty domain 
must counterbalance the smaller residual on the healthy domain 
and achieve a higher accuracy.
 
The best choice for all strategies 
seems to have a $n_F$ close to $k_F$,  e.g., for
a fault after $k_F= 11$ iterations and $\eta_{\super} = 5$ in
Tab. \ref{tab:ComparisonFault5}, the best choice is $n_I = 2$ and $n_F =10$.
The later the fault occurs, the more powerful the recovery strategy has to be.
In particular without a significantly increased compute power per degree of freedom in the faulty domain compared to the healthy domain, we cannot  fully compensate for the fault.

\section{Parallel recovery}\label{sec:parallelrecovery}
%
Now we investigate both global recovery strategies in a
parallel setting on a state-of-the-art peta-scale system.
Our test system is JUQUEEN,  
an IBM Blue Gene/Q system with a peak performance of more than $5.9$~peta\-flop/s.
Each of the 28\,672 nodes is equipped with 16 cores. 
Each core can execute up to four hardware threads
to help hiding latencies.
A five-dimensional torus network results in short communication paths within the system.
The HHG software used in this article 
is compiled by the IBM XL C/C++  compiler V12.1 
using flags \textit{-O3 -qstrict -qarch=qp -qtune=qp})
and is linked to MPICH2 version 1.5.
Technically, we realize the superman strategy  by a logical
splitting of the faulty tetrahedron and 
by employing additional MPI processes
to perform the recovery.
The software was compiled by the IBM XL C/C++  compiler V12.1 
(using flags \textit{-O3 -qstrict -qarch=qp -qtune=qp}) and linked to MPICH2 version 1.5.

Table~\ref{tab:parallelfault} presents weak scaling
results with $6 \times (2^l+1)^3$ coarse mesh tetrahedra,  $l \in \{1,2,3,4\}$.
After refinement to full resolution,
each coarse mesh tetrahedron 
contains about $2.8 \cdot 10^6$ grid points and is assigned to one hardware
thread of JUQUEEN.
Hence, the ratio between faulty and 
healthy domain decreases from $0.6\,\%$ to $3.4 \cdot 10^{-3}\,\%$ during 
the weak scaling sequence presented here.
The first column  
displays the total number of unknowns in the system.
In the largest computations, we solve for
more than $8.2 \cdot 10^{10}$ degrees of freedom on 14\,743 compute cores. 
The numbers in the rest of the table 
represent the additional time-to-solution compared to a fault-free run. 
A negative number 
here means that due to the superman strategy,
the computation with recovery terminates with the stopping criterion being satisfied
faster than a fault-free run.
As we can see from Tables~\ref{tab:parallelfault} this surprising effect is
possible due to the fact that the temporary decoupling of the subdomains reduces the  communication,
but clearly this kind of saving is not very significant.

\begin{table}[ht]
\caption[Tab.]{Additional time spans (in seconds) of different global recovery strategies during weak scaling with a fault after $k_F=7$ iterations. }
\small
  \vspace{1em} 
  \begin{tabular}{rrrrrrrrrr}
  \toprule
	    &  \multicolumn{4}{r}{DD Strategy $n_I = 1$} &   \multicolumn{4}{r}{DN Strategy $n_I =1$}\\
	     \cmidrule(r){3-6}\cmidrule(r){7-10}
  Size  & No Rec & ${\eta_{\super}}=1$&$2$&$4$&$8$&
	       ${\eta_{\super}}=1$&$2$&$4$&$8$  \\
	      \cmidrule(r){1-1}\cmidrule(r){2-2}\cmidrule(r){3-3}\cmidrule(r){4-4}\cmidrule(r){5-5}
	      \cmidrule(r){6-6}\cmidrule(r){7-7}\cmidrule(r){8-8}\cmidrule(r){9-9} \cmidrule(r){10-10}
$769^3$		&  13.00   &  12.97   &	10.30	&	2.66	&	-0.02	&	12.98	&	10.30	&	2.66	&	0.02 \\
$1\,281^3$	&  13.05   &  12.79   &	10.37	&	2.52	&	0.28	&	12.78   &	10.40	&	2.53	&	0.23 \\
$2\,305^3$	&  13.25   &  10.34   &	7.86	&	2.98	&	0.08	&	10.66	&	7.91	&	2.98	&	0.08 \\
$4\,353^3$	&  10.89   &  10.81   &	5.50	&	-0.02	&	0.15	&	10.57   &	5.29	&	-0.21	&	-0.83 \\
\bottomrule
\end{tabular}

\vspace{0.2cm}
  \begin{tabular}{rrrrrrrrrr}
  \toprule
	    &  \multicolumn{4}{r}{DD Strategy $n_I = 2$} &   \multicolumn{4}{r}{DN Strategy $n_I =2$}\\
	     \cmidrule(r){3-6}\cmidrule(r){7-10}
  Size  & No Rec & ${\eta_{\super}}=1$&$2$&$4$&$8$&
	       ${\eta_{\super}}=1$&$2$&$4$&$8$  \\
	      \cmidrule(r){1-1}\cmidrule(r){2-2}\cmidrule(r){3-3}\cmidrule(r){4-4}\cmidrule(r){5-5}
	      \cmidrule(r){6-6}\cmidrule(r){7-7}\cmidrule(r){8-8}\cmidrule(r){9-9} \cmidrule(r){10-10}
$769^3$		&		13.00			&	12.72	&	\ 5.03	&	-0.10	&	-0.24	&	12.74	&	\ 5.05	&	-0.09	&	-0.24	\\
$1\,281^3$	&		13.05			&	12.52	&	5.00	&	-0.30	&	-0.03	&	12.53	&	5.01	&	-0.28	&	-0.03	\\
$2\,305^3$	&		13.25			&	10.05	&	4.98	&	0.07	&	-0.22	&	10.11	&	5.03	&	0.07	&	-0.22	\\
$4\,353^3$	&		10.89			&	7.81	&	2.52	&	2.28	&	2.49	&	7.63	&	2.33	&	-0.54	&	-0.35	\\

\bottomrule
\end{tabular}
\vspace{0.2cm}
  \begin{tabular}{rrrrrrrrrr}
  \toprule
	    &  \multicolumn{4}{r}{DD Strategy $n_I = 3$} &   \multicolumn{4}{r}{DN Strategy $n_I =3$}\\
	     \cmidrule(r){3-6}\cmidrule(r){7-10}
  Size  &  No Rec & ${\eta_{\super}}=1$&$2$&$4$&$8$&
	       ${\eta_{\super}}=1$&$2$&$4$&$8$  \\
	      \cmidrule(r){1-1}\cmidrule(r){2-2}\cmidrule(r){3-3}\cmidrule(r){4-4}\cmidrule(r){5-5}
	      \cmidrule(r){6-6}\cmidrule(r){7-7}\cmidrule(r){8-8}\cmidrule(r){9-9} \cmidrule(r){10-10}
$769^3$	&	13.00	&	9.95		&	\ 2.28	&	-0.36	&	-0.48	
	&	9.99	&	\ 2.31	&	-0.32	&	-0.47		\\
$1\,281^3$	&	13.05	&	9.26		&	2.17	&	-0.56	&	-0.31	
	&	9.70	&	2.15	&	-0.54	&	-0.32		\\
$2\,305^3$	&	13.25	&	9.78		&	2.15	&	-0.21	&	-0.52
	&	9.81	&	2.15	&	-0.21	&	-0.51		\\
$4\,353^3$	&	10.89	&	7.53		&	2.21	&	1.99	&	2.16
	&	7.27	&	-0.65	&	-0.85	&	-0.68		\\
\bottomrule
\end{tabular}
\vspace{0.2cm}
  \begin{tabular}{rrrrrrrrrr}
  \toprule
	    &  \multicolumn{4}{r}{DD Strategy $n_I = 4$} &   \multicolumn{4}{r}{DN Strategy $n_I =4$}\\
	     \cmidrule(r){3-6}\cmidrule(r){7-10}
  Size  &  No Rec & ${\eta_{\super}}=1$&$2$&$4$&$8$&
	       ${\eta_{\super}}=1$&$2$&$4$&$8$  \\
	      \cmidrule(r){1-1}\cmidrule(r){2-2}\cmidrule(r){3-3}\cmidrule(r){4-4}\cmidrule(r){5-5}
	      \cmidrule(r){6-6}\cmidrule(r){7-7}\cmidrule(r){8-8}\cmidrule(r){9-9} \cmidrule(r){10-10}
$769^3$	&	13.00	&	9.71		&	\ 2.02	&	1.92		&	1.78	
	&	9.73	&	\ 2.04	&	1.96	&	1.80		\\
$1\,281^3$	&	13.05	&	9.43		&	1.88	&	1.71	&	1.80	
	&	9.43	&	-0.68	&	-0.82	&	-0.59		\\
$2\,305^3$	&	13.25	&	9.54		&	-0.75	&	-0.53	&	2.56
	&	9.53	&	-0.71	&	-0.51	&	-0.79		\\
$4\,353^3$	&	10.89	&	6.92		&	4.32	&	4.07	&	3.91
	&	7.15		&	1.67	&	1.47	&	1.65		\\
\bottomrule
\end{tabular}
\label{tab:parallelfault}
\end{table}

In the second column, we display the 
additional time for the do-nothing job.
The retardation of the time-to-solution ranges from $27.5 \%$ to $34.6 \%$. 
For example, 
for a large problem with $4\ 353^3$ unknowns,
the time-to-solution increases from $39.60 s$ to $50.49 s$.
To analyze the global recovery strategies, we must realize supermen with a
higher compute power
than a normal processor.
Here we implement supermen with
$\eta_{\super} \in \{1,2,4,8 \}$ by replacing the  
single faulty processor by
$\eta_{\super} $ spare processors. 
As in the previous section, the DN strategy is more robust with respect to $n_I$
compared to the DD strategy.
However, a superman power of $\eta_{\super} = 8 $ does not improve
the results significantly compared to $\eta_{\super} =4$. 
Already with $\eta_{\super} =4$,
we can in most cases fully compensate the fault and achieve the required accuracy
with no increased time-to-solution.
Setting $n_I =1$, is not a good choice since it requires a large
$\eta_{\super}$ to compensate for accuracy loss in the faulty domain.
The later the fault occurs the more it is important to set $n_I$ and thus $n_F$
large enough, i.e., re-couple not too early, and to have a strong enough superman.
We recall that $n_F = \eta_{\super} n_I$.
The DN recovery strategy in combination with $\eta_{\super} =4$, $n_I =2$ or $n_I=3$ yields 
the optimal choice with respect to time-to-solution and cost efficiency.

At this stage, we point out that the cost of the superman
becomes insignificant in a large scale computation, even if
2, 4, or 8 spare processors must be kept idle until they are called for help.
This is simply an effect of scale.
When we use 14\,743 processors, even providing 12 cores for three supermen of strength
$\eta_{\super} =4$ (which would be able to compensate for the unrealistic scenario of
three faults in a single solve)
constitutes only an extra cost of less than 0.1\%.

Finally, we test in  Tab.~\ref{tab:parallel2fault} a scenario where two faults at different locations occur at
different times. The first processor crash happens after 5 multigrid iterations and the second one after 9. 
Since we select $n_I \leq 4$, the recovery of the first fault has finished when the second one occurs.
A failure-prone cluster with many consecutive faults does not provide a suitable environment for trustable 
high-performance computations such that a certain interval between the crashes is reasonable.
For $4\ 353^3$ unknowns, the time-to-solution 
increases form $39.60 s$ to $59.34 s$ and the additional consumed computation time for the scaling ranges due to the second fault now
from $47.8 \%$ to $55.7 \%$ of the original time where no fault has occurred.

\begin{table}[ht]  
    \centering
\caption[Tab.]{Additional time spans (in seconds) of the Dirichlet-Neumann recovery strategies during weak scaling with a fault after $k_F=5$ and $k_F=9$ iterations. }
\small
  \vspace{1em} 
  \begin{tabular}{rrrrrr}
  \toprule
	  &  &  \multicolumn{4}{r}{DN Strategy $\eta_{\super} =4$}  \\
	     \cmidrule(r){3-6}
  Size  &  No Rec & ${n_I}=1$&$2$&$3$&$4$\\
	      \cmidrule(r){1-1}\cmidrule(r){2-2}\cmidrule(r){3-3}\cmidrule(r){4-4}\cmidrule(r){5-5} \cmidrule(r){6-6}
$769^3$	&	19.21	& 	8.01	&	0.05	&	-0.38	&	4.22	\\
	
$1\,281^3$&	21.27	&	10.58	&	-0.15	&	-0.68	&	3.95	\\	
		
$2\,305^3$&	18.50	&	7.91	&	-0.33	&	-0.87	&	3.76	\\
		
$4\,353^3$&	19.74	&	5.81	&	2.58	&	4.61	&	9.24	\\	
\bottomrule
\end{tabular}
\label{tab:parallel2fault}
\end{table}

We specify $\eta_{\super}=4$ and study the performance with respect to $n_I$. For $n_I=1$, we obtain already a significant improvement compared to the do-nothing job. Nevertheless with $n_I=2$ or $n_I=3$, we observe much better results.
As a rule of thumb we can set for moderate values of $n_I$, $n_I$ such that
$ n_F= \eta_{\super} n_I \approx n_I+ k_F -2$. This rule of thumb is based on the observation that we need roughly $k_F -2$ iteration to compensate for the fault in the local recovery and on the assumption that our global accuracy is still governed by the one in the faulty domain and does not suffer from the reduced communication. 
If this rule of thumb would require a value larger than three or four for $n_I$, then one should increase the
superman power. If $n_I$ is too large then too many steps without information exchange at the interface between healthy and faulty subdomain are carried out and the multigrid acts as direct solver on the subdomains with inexact boundary data. Thus short time-to-solutions require a careful balancing of volume and surface components, and a $n_I$ and $n_F$  not too small but also not too large.

\section{Conclusion and Outlook}\label{sec:conclusion}
%
This paper  
presents first insight in constructing a fault tolerant
parallel multigrid solver. 
Hard faults result in a loss of  
dynamical data in a subdomain.
Geometric multigrid solvers are 
inherently well suited to 
compensate such a loss of process state 
and to reconstruct the data based on a redundant storage scheme for numerical values 
only along the subdomain interfaces and using efficient recovery algorithms
for the bulk of the data.
To recover 
lost numerical values, local reconstruction subproblems 
with Dirichlet boundary conditions are solved approximately. 

It is found that approximate solvers using a single fine grid,
such as relaxation schemes or Krylov space methods, as well
as conventional domain decomposition techniques with direct subdomain solvers
are much too inefficient and are thus unsuitable
to serve as a basis for
practical fault recovery techniques.
In contrast, local multigrid cycles can be used to recompute the lost data with
the least numerical effort.
This strategy becomes  
efficient in both cost and time-to-solution
when the local multigrid   
recovery cycles are accelerated by a {\em superman} strategy,
as it can be realized by an excess parallelization. 

Further, we investigate  
methods 
that combine local and global processing in an asynchronous way.
A {\em Dirichlet-Dirichlet recovery} or {\em Dirichlet-Neumann recovery} strategy 
can be used,
which both iterate in the faulty subdomain and the healthy subdomain independently,
until the local solution has been recovered sufficiently well.
Only then the subdomains are reconnected to continue the regular multigrid 
solution process.
Both strategies 
can further reduce the delay in case of a fault. 
For different fault scenarios, we observe that the Dirichlet-Neumann strategy 
should be preferred.
Combined with the superman compute power on the small faulty 
subdomain, 
the global recovery techniques can result if a full numerical compensation of the fault
while costing no additional compute time.
The robustness and flexibility of the designed algorithms are tested on a  
state-of-the art peta-scale system including large scale simulation 
with close to $10^{11}$ unknowns.
We thus believe that the superman strategy
will be a viable and resource efficient approach to achieve ABFT on future exa-scale systems
which may have many millions of cores.

\cleardoublepage
\section*{Acknowledgements}
{\small This work was supported (in part) by the German Research Foundation (DFG) through the Priority Programme 1648 ``Software for Exascale Computing'' (SPPEXA). 
The authors gratefully acknowledge the Gauss Centre for Supercomputing (GCS) for providing computing time through the John von Neumann Institute for Computing (NIC) on the GCS share of the supercomputer JUQUEEN at J\"ulich Supercomputing Centre (JSC). 
The authors additionally acknowledge support by the Institute of Mathematical Sciences of the National University of Singapore, where part of this work was performed.}

\bibliographystyle{abbrv}
\bibliography{resilience}

\end{document}